# Imaging structural transitions in organometallic molecules on Ag(100) for solar thermal energy storage

Jongweon Cho[1,2,*], Ivan V. Pechenezhskiy[1,3,§], Luis Berbil-Bautista[1,3], Steven K. Meier[4], K. Peter C. Vollhardt[4], and Michael F. Crommie[1,3,*]

[1]Department of Physics, University of California at Berkeley, Berkeley, California 94720, USA

[2]Department of Physics, Myongji University, Yongin 17058, Korea

[3]Materials Sciences Division, Lawrence Berkeley National Laboratory, Berkeley, California 94720, USA

[4]Department of Chemistry, University of California at Berkeley, Berkeley, California 94720, USA

*Address correspondence to jwcho@mju.ac.kr, crommie@berkeley.edu

§Present address: Department of Physics, University of Wisconsin, Madison, Wisconsin 53706, USA

**Abstract**

The use of opto-thermal molecular energy storage at the nanoscale creates new opportunities for powering future microdevices with flexible synthetic tailorability. Practical application of these molecular materials, however, requires a deeper microscopic understanding of how their behavior is altered by the presence of different types of substrates. Here we present single-molecule resolved scanning tunneling microscopy imaging of thermally- and optically-induced structural transitions in (fulvalene)tetracarbonyldiruthenium molecules adsorbed onto a Ag(100) surface as a prototype system. Both the parent complex and the photoisomer display distinct thermally-driven phase transformations when they are in contact with a Ag(100) surface. This behavior is consistent with the loss of carbonyl ligands due to strong molecule-surface coupling. Ultraviolet radiation induces marked structural changes only in the intact parent complex, thus indicating a photoisomerization reaction. These results demonstrate how stimuli-induced structural transitions in this class of molecule depend on the nature of the underlying substrate.

**Introduction**

The ability to efficiently store energy in an easily converted form is one of the great energy challenges of our time [1]. Photo-induced structural rearrangements in organometallic molecular systems can store optical energy in covalent bonds with a high enthalpy change, offering advantages in reversibility as well as the promise of chemical tunability for tailoring optical, electronic, and steric properties for increased efficiency and energy density [2-4]. (Fulvalene)tetracarbonyldiruthenium [$FvRu_2(CO)_4$] is a robust organometallic molecule that converts light energy to chemical energy through photoisomerization, *i.e.*, the light-activated structural change of the parent complex to a high-energy isomer (see Figure 1) [5]. Energy stored in the metastable photoisomer can later be released as a measurable temperature rise in the reverse reaction upon thermal excitation or through catalysis [5-9]. However, practical device applications based on such energy storing organometallic molecules require additional understanding of how they behave in contact with conducting surfaces and under different thermodynamic environments at the single molecule level. Scanning tunneling microscopy is a well-established technique for investigating surface-bound organometallic molecules and for providing a better understanding of their behavior at the nanometer scale [10-12].

A previous STM study reported structural transitions induced by temperature and light in $FvRu_2(CO)_4$ molecules adsorbed onto Au(111) [12]. Ag(100), on the other hand, chemically interacts more strongly with most molecular adsorbates compared to Au(111) [13] and also provides a different lattice symmetry. Ag(100) has a square lattice with no large-scale surface reconstruction [14], unlike the close-packed Au(111) surface whose herringbone reconstruction [15] can substantially influence molecular self-assembly properties [12]. It is therefore useful to assess how $FvRu_2(CO)_4$ molecules behave when they are on Ag(100) compared to Au(111).

Here we report the use of cryogenic ultrahigh vacuum (UHV) scanning tunneling microscopy (STM) to investigate local thermal and photo-induced structural transitions in $FvRu_2(CO)_4$ molecules adsorbed onto Ag(100). We find that both the parent complex and the photoisomer exhibit different temperature-dependent structural phases on the Ag(100) surface. Intramolecular features revealed by STM topographic images suggest that thermally-induced CO loss takes place for both the parent complex and photoisomers in the first adsorbate layer on Ag(100), while the second molecular layer remains intact. UV light exposure results in photo-induced structural changes only for the parent complex, indicating that the parent complex photoisomerizes on Ag(100).

**Experimental Methods**

Our measurements were performed using a homebuilt variable-temperature UHV STM with optical access to the surface being probed. An atomically clean Ag(100) surface was obtained by repeated cycles of $Ar^+$ sputtering followed by thermal annealing. The synthesis of $FvRu_2(CO)_4$ and its photoisomer was conducted according to the description in Refs. 5, 16. Both were deposited *via* thermal evaporation onto clean Ag(100) substrates held at T = 12 K (NMR analysis was used to check that the process of vacuum sublimation converts only a negligible fraction of photoisomers back into the parent complex). Two different structural phases were observed following thermal treatment, referred to here as the low temperature phase (Phases 1 and 1*, where the asterisk denotes the photoisomer) and the high temperature phase (Phases 2 and 2*). The low temperature phase was acquired by removing the sample from the cryogenic STM stage using a room temperature manipulator and holding it for only 10 min before re-cooling for subsequent measurements at T = 12 K (the sample temperature is estimated to rise from 12 K to 260 ± 30 K during this 10 min interval). The high temperature phase was obtained by holding the sample at room temperature for 1.5 hr, during which time it was able to reach room temperature equilibrium. STM tunnel currents were kept below ~30 pA for stable imaging. A cw diode laser aligned at an external viewport provided UV radiation (375 nm) at the sample surface with average intensity of ~90 mW/cm$^2$. During light exposure the STM tip was retracted several millimeters from the surface and the sample temperature was maintained between 12 and 15 K.

**Results and Discussion**

When the parent complex [FvRu$_2$(CO)$_4$-PC] (Figure 1) is deposited onto clean Ag(100) in sub-monolayer amounts it is seen to assemble into two-dimensional molecular islands of the low-temperature phase (Phase 1), as shown in Figure 2a. The size of an individual molecule can be readily identified from isolated vacancies in the STM topographic images. A single molecule in Phase 1 exhibits a pair of symmetric lobes with an overall length close to the expected length of a FvRu$_2$(CO)$_4$ molecule (~8 Å).

Figure 2b shows an image of the high temperature phase (Phase 2) obtained by subjecting Phase 1 to extended annealing at room temperature. Here individual single molecules appear as single elongated lobes, in sharp contrast to the double-lobe structure of Phase 1. The long-axis dimension of a Phase 2 molecule matches the expected length of a FvRu$_2$(CO)$_4$ moiety. Phases 1 and 2 were never seen to coexist on the Ag(100) surface in the submonolayer regime.

To gain a more comprehensive understanding of the surface behavior of FvRu$_2$(CO)$_4$, we also examined its photoisomer [FvRu$_2$(CO)$_4$-PI] on Ag(100) under identical conditions as the parent complex. FvRu$_2$(CO)$_4$-PI is also seen to assemble into two structural phases on the Ag(100) surface: the low-temperature phase (Phase 1*) and the high-temperature phase (Phase 2*). Figure 3a shows a two-dimensional island of Phase 1*. Here individual molecules appear different from Phase 1 and exhibit some asymmetry around their short axis. The intramolecular contrast observed in Phase 1* suggests a nonplanar adsorption geometry.

When Phase 1* is annealed longer at room temperature, the photoisomer molecules significantly change their microscopic conformation and develop into a new high-temperature phase (Phase 2*) within the first layer resting on Ag(100) (Figure 3b). The length of individual molecules in photoisomer Phase 2* is similar to molecules in Phase 2 of the parent complex, but Phase 2* molecules have a slightly more asymmetrical (egg-like) shape than Phase 2. The second layer of the high temperature phase, on the other hand, reveals a molecular structure almost identical to the structure observed in Phase 1* (Figure 3b).

In order to examine the photoswitching activity of $FvRu_2(CO)_4$ on Ag(100) we illuminated each of the four structural phases (parent complex and photoisomer) with UV light. We observe that only Phase 1 of the parent complex exhibits any photo-induced structural changes, while the other three phases remain unperturbed by UV irradiation. Figure 4 displays a typical image of the conformational changes observed in islands of Phase 1 of the parent complex after 10 hrs of UV (375 nm) irradiation at low temperature (12 K < T < 15 K). Pronounced regions of disorder are seen to emerge within the Phase 1 islands in response to UV irradiation. The degree of disorder in Phase 1 islands is seen to monotonically increase with increased UV exposure.

We now discuss the origins of the molecular structural changes observed here in response to thermal and optical stimulation. We start with the irreversible phase transformation of the $FvRu_2(CO)_4$ parent complex from Phase 1 to Phase 2 upon extended thermal annealing at room temperature. The STM images shown in Figure 2 support a scenario in which the parent complex loses its carbonyl groups in response to extended room temperature thermal annealing. CO removal from the molecules as they switch from Phase 1 to Phase 2 exposes the Ru atoms and enhances their ability to bond with the Ag(100) surface [17]. This is similar to the thermally-induced carbonyl loss mechanism observed previously for parent complex $FvRu_2(CO)_4$ molecules deposited atop Au(111) and subjected to comparable annealing at room temperature [12].

A marked difference between the surface morphologies of Phase 2 parent complex molecules on Ag(100) and Au(111), however, is the presence of axial symmetry in the elongated molecular shape observed on Ag(100) compared to the "bent lima bean" shape observed on Au(111) [12]. Our findings imply that although the thermally-induced CO loss mechanism is at play for both molecule-decorated surfaces, it likely leads to different local adsorption geometries. Density functional theory calculations performed for the parent complex on Au(111) suggested that CO removal induces a conformation in which molecules are tilted side-ways so that the Ru atoms and cyclopentadienyl (Cp) rings are both in contact with the surface, thus leading to the asymmetric lima bean shape [12]. Ag(100), on the other hand, should have a stronger molecule-surface interaction strength compared to Au(111) [13], indicating that the symmetric appearance of Phase 2 molecules may arise from a preference for the Ru atoms to point down toward the more reactive Ag surface (Figure 2b inset). This would lead to the more symmetric morphology seen in Figure 2b.

The photoisomer of $FvRu_2(CO)_4$ also undergoes a loss of CO groups as it switches from Phase 1* to Phase 2* upon extended room temperature annealing. The analogous behavior compared to that of the parent complex likely arises from the similar CO-Ru bond strengths that exist for the two isomers [5]. This is consistent with our observation that Phase 2* molecules of the photoisomer exhibit a slightly different shape (more egg-like) compared to Phase 2 molecules (Figure 3b first layer vs. Figure 2b), likely due to the asymmetrical arrangement of Ru atoms and Cp rings (Figure 3b inset). Additional insight into the mechanism for the thermally driven Phase 1*/Phase 2* structural change can be inferred from the double layer structure of the photoisomer in Figure 3b. Here only the first layer molecules undergo a thermally-activated structural transition, whereas the second layer molecules remain unchanged. This supports an interpretation that involves Ag surface atoms mediating a structural transition.

We now consider the UV-induced structural transition observed for Phase 1 of the parent complex on Ag(100) (Figure 4). The structural disorder induced by irradiation of Phase 1 with UV light is indicative of reorganization of the parent complex into the photoisomer structure. The disordered height observed in the STM images post irradiation is consistent with a random reaction of parent complex molecules within the island structure. Photoisomerization of the parent complex (Phase 1) is a known transformation during which dinuclear metal-metal bonded carbonyls undergo metal-metal bond cleavage upon UV irradiation [4,7]. This mechanism is not operative for parent complex-based Phase 2 molecules whose properties are modified by loss of the carbonyl groups and enhanced bonding to the surface [18]. The low yield of the photo-induced structural change in Phase 1 suggests a low photoswitching efficiency for this molecule at a metallic surface compared to the higher switching efficiency seen when the molecule is in a solution environment (the surface-based switching rate is approximately one fifth of the solution-based rate [5]). This can be attributed to the proximity of the Ag surface which causes steric hindrance (FvRu$_2$(CO)$_4$ in the solid or melt is inert to photoisomerization [8] and the quantum yield for photoisomerization is qualitatively inversely proportional to solvent viscosity [9]) and leads to reduced molecular optical absorption as well as reduced excited state lifetimes [19,20].

**Conclusions**

We have investigated the self-assembly and photoswitching properties of both the parent complex and photoisomer of FvRu$_2$(CO)$_4$ adsorbed onto Ag(100) in the ultra-thin film regime. Observed thermally-induced irreversible phase transformation is attributed to CO loss from both structures based on our STM topographic images. UV-induced structural change is observed only in the low-temperature phase of the parent complex and is attributed to photoisomerization on Ag(100). Differences in molecular structure for molecules adsorbed on Ag(100) vs. Au(111) likely arise from the increased chemical reactivity of Ag(100) and resultant change in molecular orientation. The presence of a metallic substrate is seen to strongly influence the thermal- and photo-induced behavior of FvRu$_2$(CO)$_4$ and thus its energy-storage potential in the ultra-thin film limit.

**Acknowledgements**

This research was supported by the Director, Office of Science, Office of Basic Energy Sciences, Division of Materials Sciences and Engineering Division, U.S. Department of Energy under the Nanomachine program, contract no. DE-AC03-76SF0098 (molecular deposition and STM imaging), by the Sustainable Products and Solutions Program at UC Berkeley (design of molecular compounds), and by the National Science Foundation under award no. CHE-0907800 (synthesis of molecular compounds). J.C. acknowledges support from the 2015 Research Fund of Myongji University (data analysis).

**Figure captions**

**Figure 1:** Schematic for solar thermal energy storage based on FvRu$_2$(CO)$_4$ molecules. UV light from the sun induces isomerization of the parent complex to a high-energy photoisomer. The energy stored in the photoisomer can be released as heat through a thermal reversal process.

**Figure 2:** STM constant-current images of FvRu$_2$(CO)$_4$ parent complex adsorbed on a Ag(100) surface (a) before (Phase 1) and (b) after (Phase 2) extended thermal annealing at room temperature. Images acquired with $I = 30$ pA and $V_{sample} = +1$ V at $T = 12$ K. Individual molecules are outlined with blue borders. Insets show proposed structural models.

**Figure 3:** STM constant-current images of the photoisomers of FvRu$_2$(CO)$_4$ deposited onto a Ag(100) surface (a) before (Phase 1*) and (b) after (Phase 2*) extended thermal annealing at room temperature. The second layer region in (b) exhibits the same molecular morphology as Phase 1*. Images acquired with $I = 20$ pA and $V_{sample} = +1$ V at $T = 12$ K. Bottom right inset in (a) shows a close-up image of Phase 1*. Individual molecules are outlined with red borders. Insets show proposed structural models.

**Figure 4:** STM constant-current image of FvRu$_2$(CO)$_4$ parent complex (Phase 1) on Ag(100) after 10 hrs of UV light irradiation ($\lambda = 375$ nm) at low temperature (12K < T < 15K). Images acquired with $I = 30$ pA and $V_{sample} = +1$ V at $T = 12$ K. Photo-induced structural disorder seen within the molecular islands corresponds to ~10% switching yield.

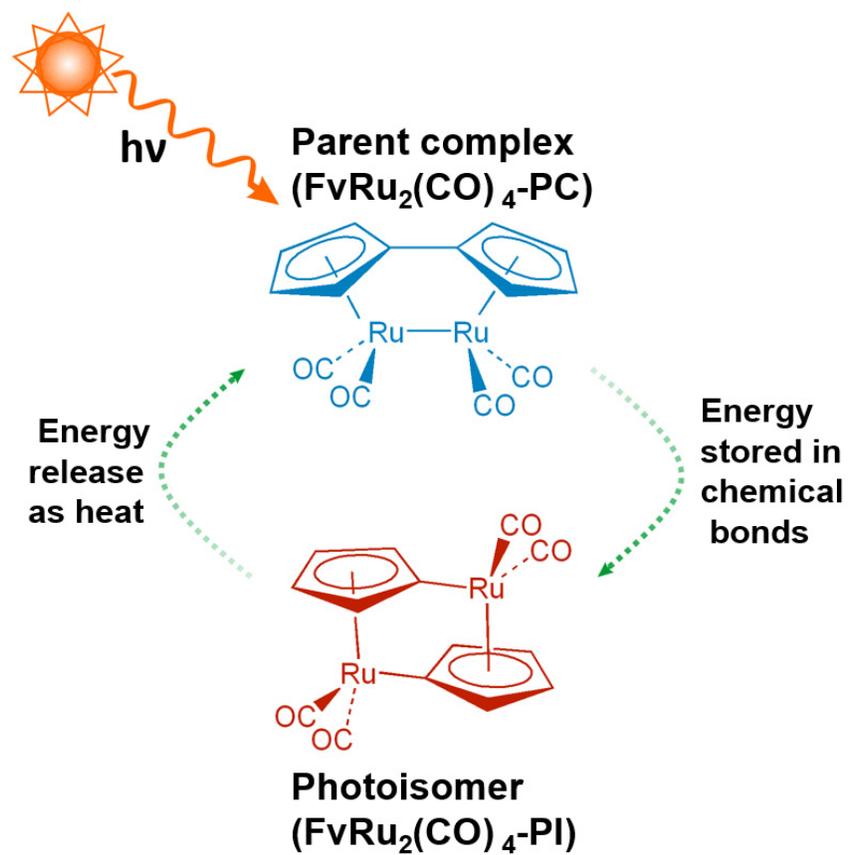

**Figure 1:**

Cho, Pechenezhskiy, Berbil-Bautista, Meier, Vollhardt, Crommie

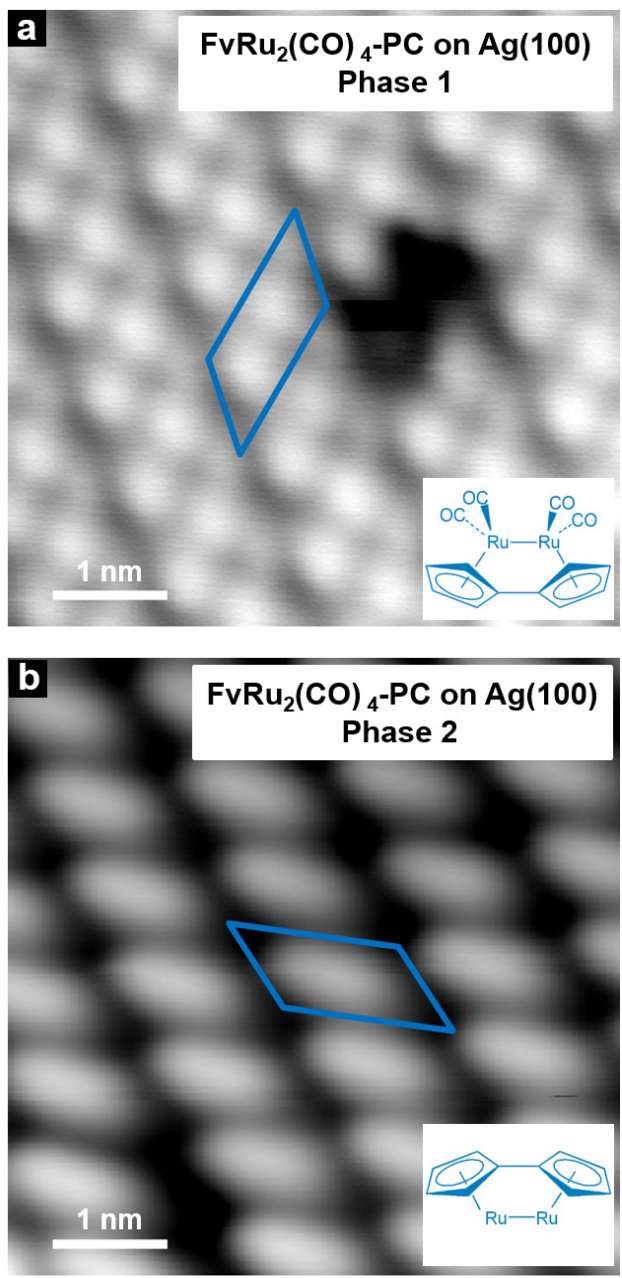

**Figure 2:**

Cho, Pechenezhskiy, Berbil-Bautista, Meier, Vollhardt, Crommie

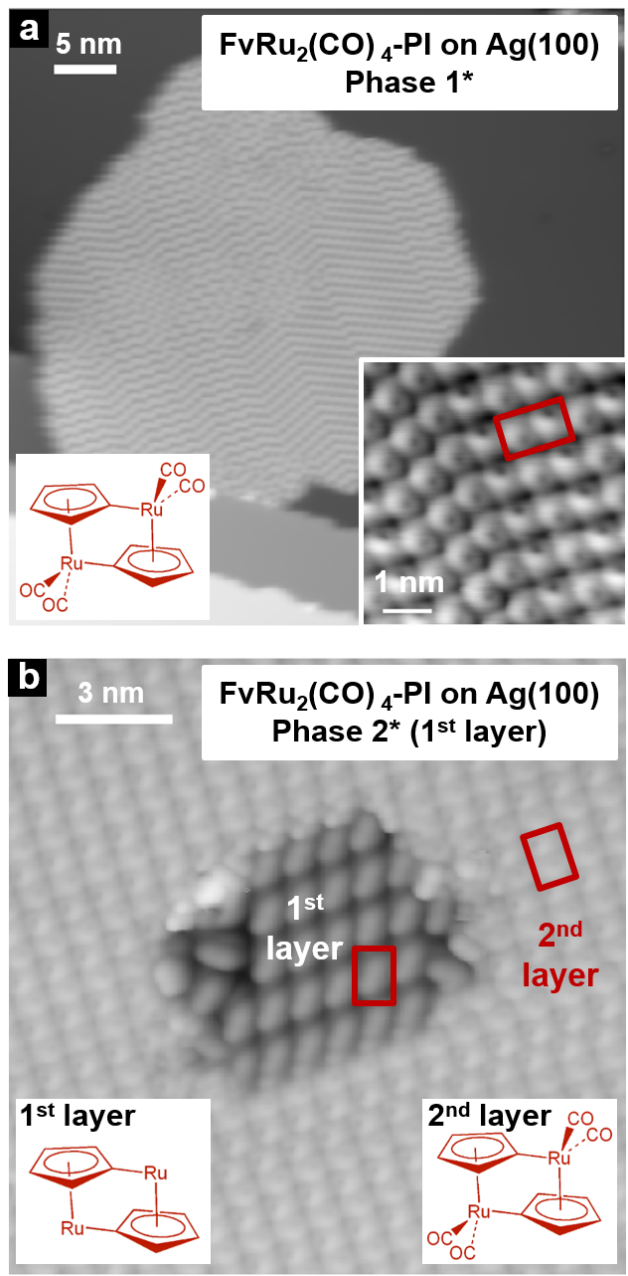

**Figure 3:**

Cho, Pechenezhskiy, Berbil-Bautista, Meier, Vollhardt, Crommie

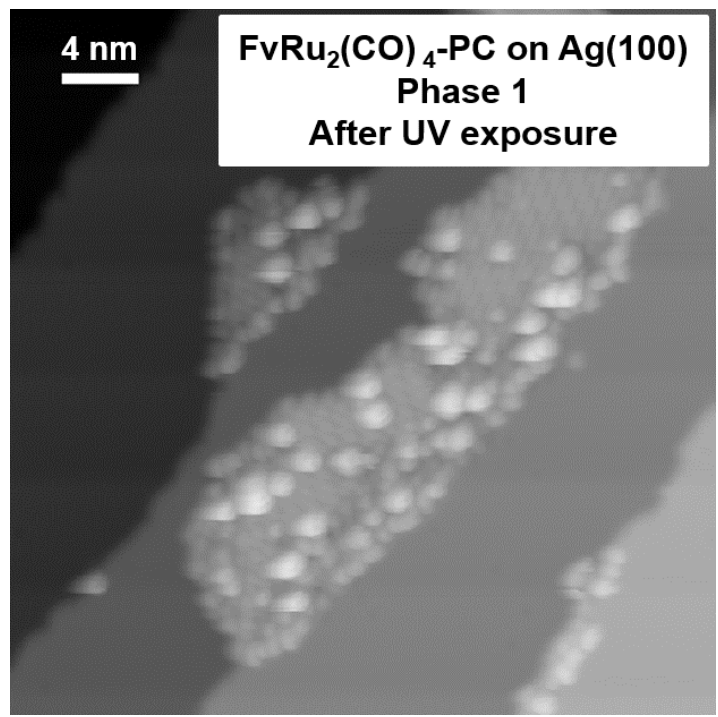

**Figure 4:**

Cho, Pechenezhskiy, Berbil-Bautista, Meier, Vollhardt, Crommie